# Wind Turbines Partial Load Power Regulation Using a Fast MPC Approach

Kamyar Ghanbarpour, Farhad Bayat, Abolfazl Jalilvand
Department of Electrical Engineering
University of Zanjan
Zanjan, Iran
Email: k.ghanbarpour@znu.ac.ir

*Abstract*—In this paper, the highly acknowledged advantages of the Model Predictive Control (MPC) approach are utilized to regulate the wind turbines' output power in the partial load region. In this region, the purpose of the designed controller is to capture maximum power from the wind. When the wind speed is above rated wind speed (full load region), control strategy is to keep output power and generator speed at their rated value. Because of the wind turbine's nonlinearities and constraints of the variables this is a challenging problem and two MPC-based methods are proposed, i.e. online and offline MPC schemes, that guarantee all constraints satisfaction and handling the systems nonlinearities. At the end, effectiveness of these two methods are compared.

## I. INTRODUCTION

Evaluation of technology has increased demands for using the modern electrical equipment. On the other hand, fossil fuels increase greenhouse gases in the atmosphere. So it is necessary to focus on the renewable energy sources available nowadays. Among the current renewable energy sources, wind energy is the world's fastest growing power [1].

The role of control system in wind turbines is very important. Wind turbine controller has different objectives according to wind speed. Above the rated power (full load region), the objective of controller is to reduce fluctuations of electrical power and rotor speed while minimizing the control actuating loads. This objective is obtained by pitching the blades to adjust the efficiency of the rotor, while applying a constant generator torque [2]. The controller that presented in this paper is designed for the wind turbines operating at low wind speeds (partial load region). The main objective of the control system in this area consists optimizing the extracted aerodynamic power. Capturing maximum aerodynamic power of wind turbine is achieved while rotor tracks optimal rotor speed of the wind turbine [3].

There are some challenges in the wind turbine control system design. Wind turbine system is a multiple-input, multiple-output (MIMO) system and there are nonlinear terms on this system [4]. Also there are some physical constraints so that control system should be able to guaranties fulfilment of all constraints [5].

Many techniques have been used to control wind turbine systems in the partial and full load region. At first, linear and classical control was designed for wind turbine described in [6-8] Because of system nonlinearity, nonlinear control methods are used in wind turbine system [9-11]. There are uncertainties on the parameters of wind turbine system, therefore robust control, sliding mode control and adaptive control methods are described in [12-15] for system robustness despite uncertainties. Also gain scheduling and LMI-based approaches are proposed in [16, 17] by choosing several operating points. One important challenge in the method of [16] is switching between the controllers which has been efficiently fixed in [17] by introducing a convex combination LMI-based approach.

The mentioned articles do not take into account the constraints and limitations on the parameters of wind turbine. Model Predictive Control is a control algorithm that can be used for the constrained systems. Multiple model predictive control and fuzzy model predictive control are proposed in [18-21]. Also the model predictive control for the wind turbine using LIDAR is designed in [22]. The other challenge in the wind turbine control system design is to cope with the nonlinearity and taking into account the constraints. So an approach is proposed in this paper for overcoming the mentioned challenges.

The objective of this paper is to design a controller for capturing maximum power in the partial load region that takes into consideration the nonlinear nature of the wind turbine behavior and constraints on the wind turbine parameters. Two strategies are used to cope with challenges that mentioned above. At first, two operation points in the partial load region are considered and wind turbine system is linearized at these points. Then offline MPC problem is solved for each of the points. At second strategy, wind turbine system is linearized at every sampling time and online MPC problem is solved at each sampling time. Control inputs are obtained at every sampling time by solving online MPC and can be applied on wind turbine system.

## II. WIND TURBINE MODELLING

A wind turbine is modeled as an interconnection of several subsystems including: aerodynamics, drive train, pith control, generator and converter (Fig. 1).

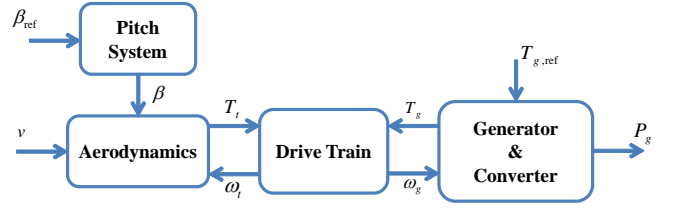

Fig. 1. Wind turbine system model

### A. Aerodynamic Model

Aerodynamic model describes how the wind power is converted to the rotational energy. The aerodynamic power captured by the rotor is given by:

$$P_t(t) = \frac{1}{2}\rho\pi R^2 v^3(t) C_p(\lambda(t), \beta(t)) \qquad (1)$$

where $\rho$ is air density, $R$ is blade length, $v$ is effective wind speed and $C_p$ is the power coefficient that depends on the blade pitch angle $\beta$ and Tip-Speed-Ratio $\lambda$. The power coefficient $C_p$ is given by (see Fig. 2):

$$C_p(\lambda, \beta) = 0.5176\left(\frac{116}{\lambda_i} - 0.4\beta - 5\right)e^{-\frac{21}{\lambda_i}} + 0.0068\lambda \qquad (2)$$

$$\frac{1}{\lambda_i} = \frac{1}{\lambda + 0.08\beta} - \frac{0.035}{\beta^3 + 1} \qquad (3)$$

$$\lambda(t) = \frac{\omega_t(t)R}{v(t)} \qquad (4)$$

Then, the aerodynamic torque is obtained as:

$$T_t(t) = \frac{P_t(t)}{\omega_t(t)} = \frac{C_p(\lambda(t), \beta(t))}{\lambda(t)}\frac{1}{2}\rho\pi R^3 v^2(t) \qquad (5)$$

### B. Drive Train Model

The drive train converts high torque on the low speed shaft to low torque on the high speed shaft to provide the speed requirement of the generator. The shafts are interconnected by a transmission line with a gear ratio $N_g$ modeled as:

$$J_t \dot{\omega}_t = T_t - N_g T_{tw} \qquad (6)$$

$$J_g \dot{\omega}_g = T_{tw} - T_g \qquad (7)$$

$$T_{tw} = K_s(N_g \theta_t - \theta_g) + B_s(N_g \omega_t - \omega_g) \qquad (8)$$

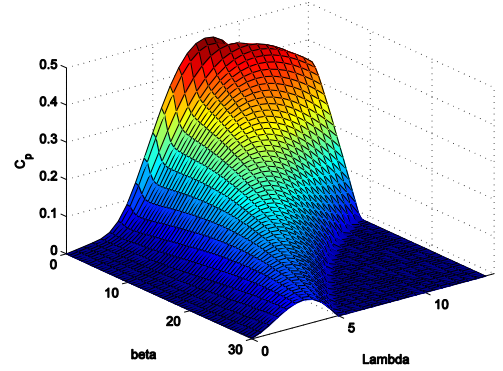

Fig. 2. $C_p(\lambda, \beta)$ curve

where $J_t$ and $J_g$ are the inertia of the turbine and generator, respectively. $T_{tw}$ is the drive train torsional torque. $N_g$ is gear ratio. $K_s$ and $B_s$ are the shaft stiffness and damping coefficients, respectively.

### C. Pitch System Model

Pitch angle control subsystem is modeled as a first-order dynamic as:

$$\dot{\beta} = -\frac{1}{\tau}\beta + \frac{1}{\tau}\beta_{ref} \qquad (9)$$

where $\beta$ and $\beta_{ref}$ are the pitch angle and reference input of pitch angle, respectively. $\tau$ is the time constant. It is important to note that the following constraints need to be considered:

$$\beta_{min} \leq \beta \leq \beta_{max} \quad , \quad \dot{\beta}_{min} \leq \dot{\beta} \leq \dot{\beta}_{max} \qquad (10)$$

### D. Generator and Converter Models

In the wind turbine, since the dynamics of the electrical subsystems are much faster than the turbine dynamics then this subsystem can be modeled as a first-order model:

$$\dot{T}_g = -\frac{1}{\tau_g}T_g + \frac{1}{\tau_g}T_{g,ref} \qquad (11)$$

$$P_g = T_g \omega_g \eta \qquad (12)$$

where $T_g$ and $T_{g,ref}$ are the generator torque and reference input of generator torque, respectively. $\eta$ is efficiency, $\tau_g$ is time constant and $P_g$ is the generator power.

*E. Unified Model*

integrating all discussed subsystems, the unified model of the wind turbine is achieved:

$$\dot{x}(t) = Ax(t) + Bu(t) + f(x,v) \quad (13)$$

$$y(t) = \begin{bmatrix} \omega_g & P_g \end{bmatrix}^T \quad (14)$$

where $x = \begin{bmatrix} \omega_t, \omega_g, T_{tw}, T_g, \beta \end{bmatrix}^T$ and $u = \begin{bmatrix} T_g^*, \beta^* \end{bmatrix}^T$ represent the state and input vectors, respectively. $f(x,v) = B_2 T_t$ is nonlinear part of wind turbine model, and

$$A = \begin{bmatrix} 0 & 0 & -\frac{N_g}{J_t} & 0 & 0 \\ 0 & 0 & \frac{1}{J_g} & -\frac{1}{J_g} & 0 \\ K_s N_g & -K_s & -\left(\frac{N_g^2 B_s}{J_t} + \frac{B_s}{J_g}\right) & \frac{B_s}{J_g} & 0 \\ 0 & 0 & 0 & -\frac{1}{\tau_g} & 0 \\ 0 & 0 & 0 & 0 & \frac{1}{\tau} \end{bmatrix} \quad (15\text{-a})$$

$$B = \begin{bmatrix} 0 & 0 \\ 0 & 0 \\ 0 & 0 \\ \frac{1}{\tau_g} & 0 \\ 0 & \frac{1}{\tau} \end{bmatrix}, B_2 = \begin{bmatrix} \frac{1}{J_t} \\ 0 \\ \frac{N_g B_s}{J_t} \\ 0 \\ 0 \end{bmatrix} \quad (15\text{-b})$$

### III. CONTROL OBJECTIVES

In partial load region, i.e. under the rated wind speed, the main control objective in the wind turbine is to maximize the extracted aerodynamic power. To this aim and according to (1), the maximum aerodynamic power is obtained when $C_p(\lambda, \beta)$ is in its maximum value. As shown in Fig. 2, $C_p(\lambda, \beta)$ has a unique maximum in $\lambda_{opt}$ and $\beta_{opt}$, i.e. $C_p(\lambda_{opt}, \beta_{opt}) = C_{p_{opt}}$. Therefore, from (2), optimal rotor speed is obtained:

$$\omega_{t_{opt}} = \frac{\lambda_{opt}}{R} v \quad (16)$$

Optimal rotor speed is a reference signal of controller and should be tracked by the rotor speed of wind turbine. Also $\omega_{g_{opt}} = N_g \omega_{t_{opt}}$ at the steady state condition. So when the generator speed tracks the optimal generator speed, Maximum power capturing is achieved [10]. In the partial load region, the main challenge for the controller design is aerodynamic power maximization while minimizing transient loads. Another challenge is the nonlinearity of wind turbine system dynamics and continuous variation of operating point. Finally, as mentioned above, there are constraints on the variables of wind turbine system. The constraints in the wind turbine system that considered on this paper are amplitude and speed of the pitch actuators, the generated power and the turbine speed [18]. The control system should be able to satisfy these constraints on the acceptable limits. Proposed strategy causes to wind turbine system be robust against the mentioned challenges. Offline and online MPC is designed to guarantee all of the wind turbine system constraints and to cope with nonlinearities. Then the results of these two strategies are compared together.

### IV. MODEL PREDICTIVE CONTROL DESIGN FOR WIND TURBINE

Model predictive control (MPC) approach is acknowledged for its exceptional ability of handling constraints and achieving high performance control objectives. To this aim, first the linearized model of the wind turbine is obtained around its operating point. The nonlinearity of the wind turbine is related to the aerodynamic torque that defined in (5), so one obtains:

$$\delta T_t = \underbrace{\frac{\partial T_t}{\partial \omega_t}\bigg|_{(\bar{\omega}_t, \bar{v}, \bar{\beta})}}_{L_\omega} \delta \omega_t + \underbrace{\frac{\partial T_t}{\partial v}\bigg|_{(\bar{\omega}_t, \bar{v}, \bar{\beta})}}_{L_v} \delta v + \underbrace{\frac{\partial T_t}{\partial \beta}\bigg|_{(\bar{\omega}_t, \bar{v}, \bar{\beta})}}_{L_\beta} \delta \beta \quad (17)$$

where $\delta$ represents the deviation of variable from its operating point. $\bar{\omega}_t$ and $\bar{\beta}$ are dependent on $\bar{v}$ that measured by anemometer in every sampling time. Continuous-time linearized state space model is defined as:

$$\begin{aligned} \delta \dot{x}(t) &= A_c \delta x(t) + B_{cu} \delta u(t) + B_{cv} \delta v(t) \\ \delta y(t) &= C_c \delta x(t) \end{aligned} \quad (18)$$

where

$$A_c = \begin{bmatrix} \frac{L_\omega}{J_t} & 0 & -\frac{N_g}{J_t} & 0 & \frac{L_\beta}{J_t} \\ 0 & 0 & \frac{1}{J_g} & -\frac{1}{J_g} & 0 \\ \Phi & -K_s & \Psi & \frac{B_s}{J_g} & \frac{N_g B_s}{J_t} L_\beta \\ 0 & 0 & 0 & -\frac{1}{\tau_g} & 0 \\ 0 & 0 & 0 & 0 & -\frac{1}{\tau} \end{bmatrix} \quad (19)$$

$$B_{cu} = \begin{bmatrix} 0 & 0 \\ 0 & 0 \\ 0 & 0 \\ \frac{1}{\tau_g} & 0 \\ 0 & \frac{1}{\tau} \end{bmatrix}, B_{cv} = \begin{bmatrix} \frac{L_v}{J_t} \\ 0 \\ \frac{N_g B_s}{J_t} L_v \\ 0 \\ 0 \end{bmatrix} \quad (20)$$

$$C_c = \begin{bmatrix} 0 & 1 & 0 & 0 & 0 \\ 0 & \eta \bar{T}_g & 0 & \eta \bar{\omega}_g & 0 \end{bmatrix} \quad (21)$$

where $\Phi = K_s N_g + \frac{N_g B_s}{J_t} L_\omega$, and $\Psi = -\left(\frac{N_g^2 B_s}{J_t} + \frac{B_s}{J_g}\right)$.

By discretizing (18) at sampling time $T_s$, one obtains:

$$\begin{aligned} \delta x(k+1) &= A_D \delta x(k) + B_{Du} \delta u(k) + B_d d(k) \\ \delta y(k) &= C_D \delta x(k) \end{aligned} \quad (22)$$

where $\delta u(k)$ is control input vector, $\delta x(k)$ is state vector and $\delta y(k)$ is controlled measurement vector and at every sampling time are defined as bellow:

$$\begin{aligned} \delta x(k) &\stackrel{\text{def}}{=} \begin{bmatrix} \delta \omega_t(k) & \delta \omega_g(k) & \delta T_{tw}(k) & \delta T_g(k) & \delta \beta(k) \end{bmatrix}^T \\ \delta u(k) &\stackrel{\text{def}}{=} \begin{bmatrix} \delta T_{g,\text{ref}}(k) & \delta \beta_{\text{ref}}(k) \end{bmatrix}^T \\ \delta y(k) &\stackrel{\text{def}}{=} \begin{bmatrix} \delta \omega_g(k) & \delta P_g(k) \end{bmatrix}^T \end{aligned} \quad (23)$$

The matrices $A_D$ and $B_{Du}$ are calculated from continuous-time matrices defined in (19)-(21) as:

$$A_D = e^{A_c T_s}, B_{Du} = \int_0^{T_s} B_{cu} e^{A_c \tau} d\tau \quad (24)$$

The effects of unknown but constant mean disturbance are represented by $d(k)$ with the following dynamics:

$$d(k+1) = d(k) \quad (25)$$

Combining (25) and (22), the augmented prediction model used in the MPC formulation is given by:

$$\begin{aligned} \begin{bmatrix} \delta x(k+1) \\ d^i(k+1) \end{bmatrix} &= \begin{bmatrix} A_D & B_d \\ 0 & I \end{bmatrix} \begin{bmatrix} \delta x(k) \\ d(k) \end{bmatrix} + \begin{bmatrix} B_{Du} \\ 0 \end{bmatrix} \delta u(k) \\ \delta y(k) &= \begin{bmatrix} C_D & 0 \end{bmatrix} \begin{bmatrix} \delta x(k) \\ d(k) \end{bmatrix} \end{aligned} \quad (26)$$

This can be represented by the following form:

$$\begin{aligned} x_p(k+1) &= A_p x_p(k) + B_p u_p(k) \\ y_p(k) &= C_p x_p(k) + D_p u_p(k) \end{aligned} \quad (27)$$

A. *Designing MPC for wind turbine*

In order to achieve an offset-free tracking MPC-based controller, an integral action is required to be added by representing the model in terms of the control move $\Delta u(k) = u_p(k) - u_p(k-1)$. A new state $x_u(k)$ is defined as $x_u(k) = u_p(k-1)$, then (27) can be augmented as:

$$\begin{aligned} \begin{bmatrix} x_p(k+1) \\ x_u(k+1) \end{bmatrix} &= \begin{bmatrix} A_p & B_p \\ 0 & I \end{bmatrix} \begin{bmatrix} x_p(k) \\ x_u(k) \end{bmatrix} + \begin{bmatrix} B_p \\ I \end{bmatrix} \Delta u(k) \\ y_p(k) &= \begin{bmatrix} C_p & 0 \end{bmatrix} \begin{bmatrix} x_p(k) \\ x_u(k) \end{bmatrix} \end{aligned} \quad (28)$$

This can be equivalently written as:

$$\begin{aligned} x_a(k+1) &= A_a x_a(k) + B_a \Delta u(k) \\ y_a(k) &= C_a x_a(k) \end{aligned} \quad (29)$$

The MPC based controller solves the following constrained optimization problem at each sampling time:

$$\min_{\substack{\Delta u(k+j) \\ j=0,1,\dots,N_C-1}} \left\{ \begin{array}{l} \sum_{j=1}^{j=N_p} e^T(k+j) Q e(k+j) + \\ \sum_{j=0}^{j=N_C-1} \Delta u^T(k+j) R \Delta u(k+j) + \\ \sum_{j=0}^{j=N_C-1} u^T(k+j) R_u u(k+j) \end{array} \right\} \quad (30)$$

subject to:

$$\begin{aligned} &x_a(k+1) = A_a x_a(k) + B_a \Delta u(k) \\ &y_a(k) = C_a x_a(k) \\ &\Delta \beta_{\min} \leq \Delta \beta_{\text{ref}}(k+j) \leq \Delta \beta_{\max}, j = 1, 2, \dots, N_c \\ &\beta_{\min} \leq \beta_{\text{ref}}(k+j) \leq \beta_{\max}, j = 1, 2, \dots, N_c \\ &0 \leq T_{g,\text{ref}} \leq T_{g,\max}, j = 1, 2, \dots, N_c \\ &\omega_g(k+j) \leq \omega_{g,\max}, j = 1, 2, \dots, N_p \\ &P_g(k+j) \leq P_{g,\max}, j = 1, 2, \dots, N_p \end{aligned} \quad (31)$$

where weighting matrices are defined as:

$$Q^i = \text{diag}(q_1, q_2), R = \text{diag}(r_1, r_2), R_u = \text{diag}(0, r_3) \quad (32)$$

**Proposition 1.** The quadratic optimization problem (30) subject to the constraints (31) can be converted to the following form:

$$\begin{aligned} &\min \frac{1}{2} \Delta U' H \Delta U + \begin{bmatrix} x'(k) & R'_s \end{bmatrix} F \Delta U \\ &\text{s.t.} G \Delta U \leq W + S \begin{bmatrix} x(k) \\ R_s(k) \end{bmatrix} \end{aligned} \quad (33)$$

*Proof.* (30) can be written as

$$Y'Q_1Y - Y'Q_1R_s - R'_sQ_1Y + R'_sQ_1R_s + \Delta U'R_1\Delta U + U'R_{u1}U \quad (34)$$

where

$$\begin{aligned}
Y &= \left[y^T(k+1), y^T(k+2), \cdots, y^T(k+N_p)\right]^T \\
R_s &= \left[r^T(k+1), r^T(k+2), \cdots, r^T(k+N_p)\right]^T \\
\Delta U &= \left[\Delta u^T(k), \Delta u^T(k+1), \cdots, \Delta u^T(k+N_c-1)\right]^T \\
U &= \left[u^T(k), u^T(k+1), \cdots, u^T(k+N_c-1)\right]^T \\
Q_1 &= diag(Q), R_1 = diag(R), R_{u1} = diag(R_u)
\end{aligned} \quad (35)$$

Furthermore, based on (29) and by definition $u_p(k) = u_p(k-1) + \Delta u(k)$, the following equations can be obtained:

$$\begin{aligned}
X &= T_1 x(k) + S_1 \Delta U \\
Y &= C_1 X = C_1 T_1 x(k) + C_1 S_1 \Delta U \\
U &= L_1 x(k) + L_2 \Delta U
\end{aligned} \quad (36)$$

where

$$X = \left[x^T(k+1), x^T(k+2), \cdots, x^T(k+N_p)\right]^T$$

$$T_1 = \begin{bmatrix} A \\ A^2 \\ \vdots \\ A^{N_p} \end{bmatrix}, S_1 = \begin{bmatrix} B & 0 & \cdots & 0 \\ AB & B & \cdots & 0 \\ \vdots & \vdots & \ddots & \vdots \\ A^{N_p-1}B & A^{N_p-2}B & \cdots & A^{N_p-N_c}B \end{bmatrix} \quad (37)$$

$$C_1 = \begin{bmatrix} C & & & \\ & C & & \\ & & \ddots & \\ & & & C \end{bmatrix}, L_1 = \begin{bmatrix} 0 & I \\ 0 & I \\ \vdots & \vdots \\ 0 & I \end{bmatrix}, L_2 = \begin{bmatrix} I & 0 & \cdots & 0 \\ I & I & \ddots & \vdots \\ \vdots & \vdots & \ddots & 0 \\ I & I & \cdots & I \end{bmatrix}$$

By replacing (36) into (34), matrices $H$ and $F$ can be obtained as:

$$H = 2S_1^T C_1^T Q_1 C_1 S_1 + 2R_1 + 2L_2^T R_{u1} L_2 \quad (38)$$

$$F = \begin{bmatrix} 2T_1^T C_1^T Q_1 C_1 S_1 + 2L_1^T R_{u1} L_2 \\ -2Q_1 C_1 S_1 \end{bmatrix} \quad (39)$$

Also the constraint of wind turbine system can be written as:

$$\begin{aligned}
\Delta u_{min} &\leq \Delta u(k+j) \leq \Delta u_{max} \\
u_{min} &\leq u_p(k+j) \leq u_{max} \\
y_{min} &\leq y_p(k+j) \leq y_{max}
\end{aligned} \quad (40)$$

Considering $U = L_1 x(k) + L_2 \Delta U$ from (36) and also

$$U_{max/min} = \begin{bmatrix} u_{max/min} \\ u_{max/min} \\ \vdots \\ u_{max/min} \end{bmatrix}, Y_{max/min} = \begin{bmatrix} y_{max/min} \\ y_{max/min} \\ \vdots \\ y_{max/min} \end{bmatrix}, \Delta U_{max/min} = \begin{bmatrix} \Delta u_{max/min} \\ \Delta u_{max/min} \\ \vdots \\ \Delta u_{max/min} \end{bmatrix} \quad (41)$$

Finally, (40) can be converted to constrain part of (33) with the following matrices:

$$G = \begin{bmatrix} C_1 S_1 \\ -C_1 S_1 \\ L_2 \\ -L_2 \\ I_N \\ -I_N \end{bmatrix}, W = \begin{bmatrix} Y_{max} \\ -Y_{min} \\ U_{max} \\ -U_{min} \\ \Delta U_{max} \\ -\Delta U_{min} \end{bmatrix}, S = \begin{bmatrix} -C_1 T_1 & 0 \\ C_1 T_1 & 0 \\ -L_1 & 0 \\ L_1 & 0 \\ 0 & 0 \\ 0 & 0 \end{bmatrix} \quad (42)$$

This completes the proof. □

### B. Online and Offline MPC

Discrete-time linearized state space model for wind turbine system given in (22). Two methods are proposed for solving MPC problem regarding to the linearization of the nonlinear model at operation point $\bar{v}$. Offline strategy is to select two operation points in the partial load region and linearize the nonlinear system at these points. One of these two linearized mode is used regarding to wind speed horizon and offline MPC problem is solved to determine control inputs. Adding operating points is able to increase the accuracy of linearized model in comparison with the nonlinear model of wind turbine. But numerous operating points causes switching problem between the extracted models. So two operating points are selected considering the results of simulation. The algorithm of offline method is shown in Fig. 3-(a).

In the second strategy, online MPC problem is solved. In this method, every sampling time is considered as operation point and linearization is performed in every sampling time. So there are $N$ linear model of wind turbine that $N$ is the number of samples. Then linear online MPC is designed for the $N$ linear model of wind turbine. The algorithm of offline method is shown in Fig. 3-(b).

**Remark 1:** The proposed offline MPC approach reduces the online computational complexity to a large extent by moving offline some computationally demanding matrix calculations in the online optimization problem. It is worth noting that the online computations can be further reduced by means of the multi-parametric programming tool [22-24].

**Remark 2:** It is noted that one way to increase the available online computation time rather than increasing the embedded system's computational power is to increase the sampling time of the system. However, the sampling time has to be properly chosen ensuring that the discrete-time controller will have favorable results when applied to the continuous-time system by capturing the complete dynamics of the continuous-time system. Otherwise, the performance of the closed-loop system will severely be degraded and even lead to infeasible and undesirable results. The problem of the analytical calculation of the maximum allowable sampling period has been recently studied in the context of sampled-data control systems (see e.g. [25, 26]).

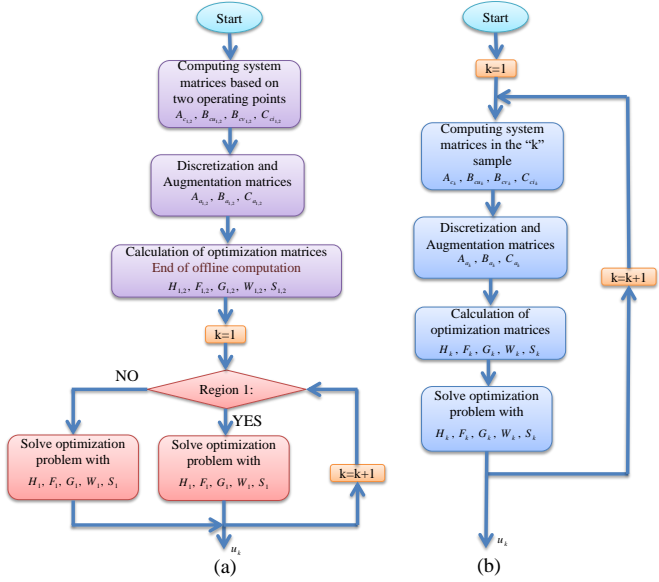

**Fig. 3. Algorithm of a) offline and b) online method**

## V. SIMULATION RESULTS

Simulation results are based on a wind turbine that its characteristics are given in Appendix A. Wind speed profile that considered in this paper is shown in Fig. 4. In this section, two cases of simulation are presented. Both of cases are based on MPC tracking problem with following weights and horizons.

$$q_1 = 100, q_2 = 0, r_1 = 10^{-6}, r_2 = 10^3, r_3 = 10^3, N_p = 20, N_c = 5$$

**Case1:** At first, offline MPC is considered for maximum power tracking. Two operating points are considered in this method. So that for the wind speed range of $v_{\Re_1} = [4, 8.7)$, operation point is 6.4 and for $v_{\Re_1} = [8.7, 11)$, operation point is 10. Two different state space matrices have been obtained and one of these models is chosen according to the wind speed. Control inputs that are obtained by offline MPC in two sub-regions, are applied to wind turbine system. Tracking of optimal generator speed and maximum power captured are shown in Fig. 5. Also Fig. 6 shows control inputs of wind turbine system. As shown in Fig. 5 and Fig. 6, generator speed tracks reference signal and control inputs constraints have been fulfilled.

**Case2:** second strategy is to obtain state space matrices at every sampling time and find the control inputs by MPC optimization problem at every sampling time. Tracking signals that are obtained by online MPC method are shown in Fig. 7. Also control inputs are shown in Fig. 8. As shown in this Fig. 8, fluctuation of generator torque in online MPC is less than offline MPC. Complexity for solving online MPC problem is more than offline method but maximum power tracking with online method is more accurate than offline MPC. In order to compare these two methods, tracking error between the maximum power and captured power of wind turbine is shown in Fig. 9. As shown in Fig. 9, tracking error of online method is less than offline method.

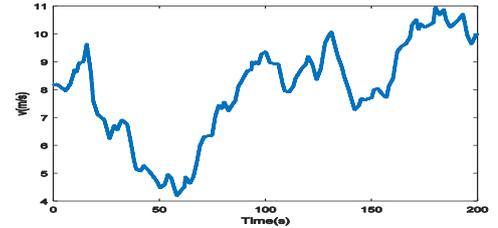

**Fig. 4. Wind speed profile**

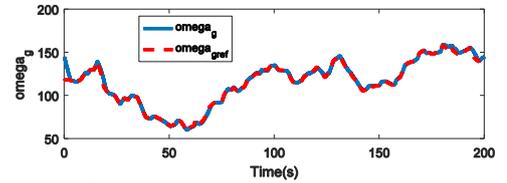

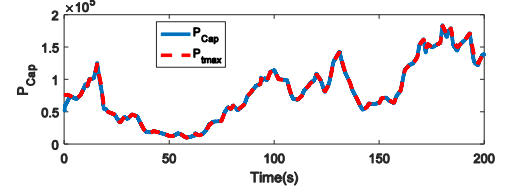

**Fig. 5. Generator speed and wind turbine power with offline MPC**

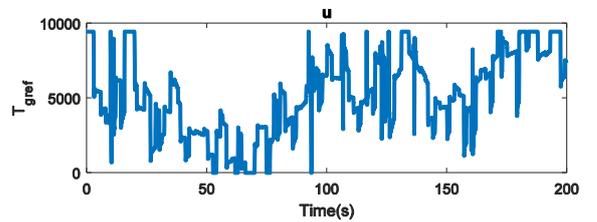

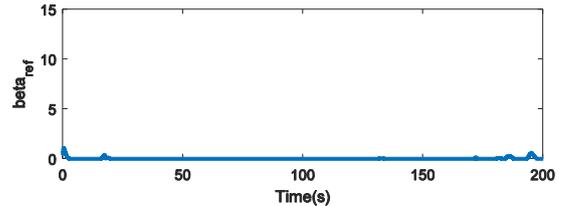

**Fig. 6. Control inputs with offline MPC**

## VI. CONCLUSION

Offline and online methods are used in this paper to solve MPC problem for capturing maximum power tracking in the wind turbine system. In offline method, two linear state space models are obtained and MPC problem are solved for each of these two models. In online method, linear state space model is obtained in every sampling time. Online MPC problem is solved regarding to obtained linear model in all of sampling time and control inputs can be obtained. Simulation results show that maximum power tracking is better than offline MPC despite sophisticated calculations. Also fluctuation of control inputs in online MPC is less than offline MPC.

## APPENDIX A: WIND TURBINE DATA

$\rho = 1.225 \text{kg/m}^3, R = 35\text{m}, J_t = 1.86 \times 10^6 \text{kgm}^2, J_g = 56.29 \text{kgm}^2,$
$N_g = 62.6, K_s = 31.8 \times 10^4, B_s = 212.2 \text{Nm/rad/s}, \tau = 0.1\text{s},$
$\tau_g = 0.02\text{s}, \beta_{min} = 0°, \beta_{max} = 45°, \dot{\beta}_{min} = -10°/s, \dot{\beta}_{max} = 10°/s,$
$\lambda_{opt} = 8.1, \beta_{opt} = 0, C_{P_{opt}} = 0.48, T_s = 50\text{ms}$

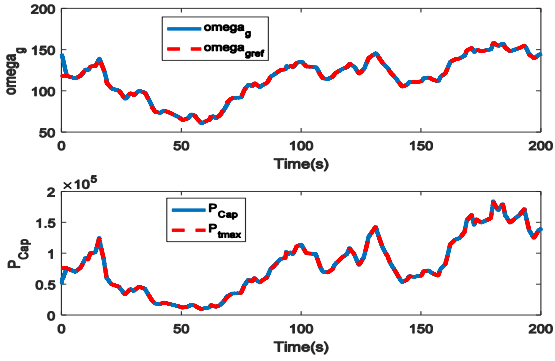

**Fig. 7. Generator speed and aerodynamic power of wind turbine with online MPC**

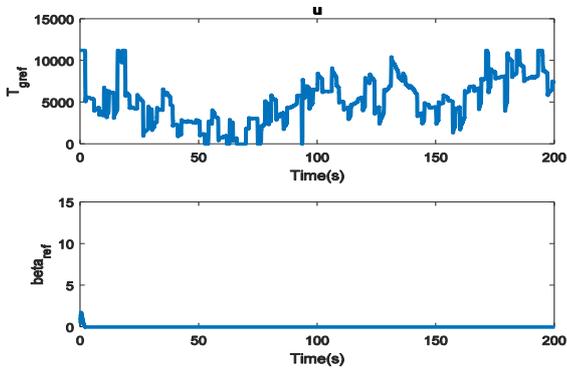

**Fig. 8. Control inputs with online MPC**

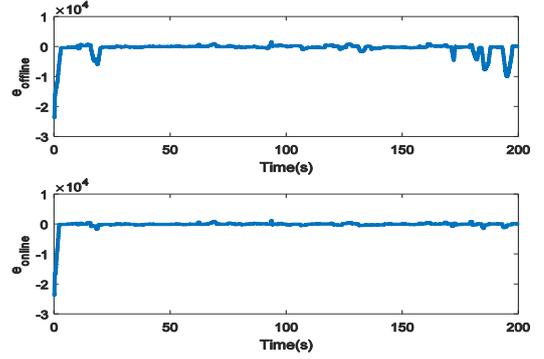

**Fig. 9. Tracking error comparison between offline and online method**

## REFERENCES

[1] G. W. E. Council, Global wind report 2010, Online: www.gwec.net/index.php.

[2] B. Boukhezzar, L. Lupu, H. Siguerdidjane, M. Hand, Multivariable control strategy for variable speed, variable pitch wind turbines, Renewable Energy 32 (8) (2007) 1273–1287.

[3] K. Z. Østergaard, J. Stoustrup, P. Brath, Linear parameter varying control of wind turbines covering both partial load and full load conditions, International Journal of Robust and Nonlinear Control 19 (1) (2009) 92–116.

[4] B. Boukhezzar, H. Siguerdidjane, Nonlinear control with wind estimation of a dfig variable speed wind turbine for power capture optimization, Energy Conversion and Management 50 (4) (2009) 885–892.

[5] D. Schlipf, D. J. Schlipf, M. Kühn, Nonlinear model predictive control of wind turbines using lidar, Wind Energy 16 (7) (2013) 1107–1129.

[6] E.A. Bossanyi, The design of closed loop controllers for wind turbines, Wind energy 3 (3) (2000) 149–163.

[7] I. Munteanu, N. A. Cutululis, A. I. Bratcu, E. Ceanga, Optimization of variable speed wind power systems based on a lqg approach, Control engineering practice 13 (7) (2005) 903–912.

[8] H. Camblong, I. M. de Alegria, M. Rodriguez, G. Abad, Experimental evaluation of wind turbines maximum power point tracking controllers, Energy Conversion and Management 47 (18-19) (2006) 2846–2858.

[9] B. Boukhezzar, H. Siguerdidjane, Comparison between linear and nonlinear control strategies for variable speed wind turbines, Control Engineering Practice 18 (12) (2010) 1357–1368.

[10] B. Boukhezzar, H. Siguerdidjane, M. M. Hand, Nonlinear control of variable-speed wind turbines for generator torque limiting and power optimization, Journal of solar energy engineering 128 (4) (2006) 516–530.

[11] E. Muhando, T. Senjyu, A. Yona, H. Kinjo, T. Funabashi, Disturbance rejection by dual pitch control and self-tuning regulator for wind turbine generator parametric uncertainty compensation, IET Control Theory & Applications 1 (5) (2007) 1431–1440.

[12] B. Beltran, M. E. H. Benbouzid, T. Ahmed-Ali, Second-order sliding mode control of a doubly fed induction generator driven wind turbine, IEEE Transactions on Energy Conversion 27 (2) (2012) 261–269.

[13] E. Kamal, A. Aitouche, R. Ghorbani, M. Bayart, Robust nonlinear control of wind energy conversion systems, International Journal of Electrical Power & Energy Systems 44 (1) (2013) 202–209.

[14] J. Mérida, L. T. Aguilar, J. Dávila, Analysis and synthesis of sliding mode control for large scale variable speed wind turbine for power optimization, Renewable Energy 71 (2014) 715–728.

[15] X.-x. Yin, Y.-g. Lin, W. Li, Y.-j. Gu, P.-f. Lei, H.-w. Liu, Adaptive backstepping pitch angle control for wind turbine based on a new electrohydraulic pitch system, International Journal of Control 88 (11) (2015) 2316–2326.

[16] F. Bianchi, R. Mantz, C. Christiansen, Control of variable-speed wind turbines by lpv gain scheduling, Wind Energy 7 (1) (2004) 1–8.

[17] F. Bayat, and H. Bahmani. Power regulation and control of wind turbines: LMI-based output feedback approach, International Transactions on Electrical Energy Systems 27 (2017) e2450.

[18] M. Soliman, O. Malik, D. Westwick, Multiple model multiple-input multiple-output predictive control for variable speed variable pitch wind energy conversion systems, IET renewable power generation 5 (2) (2011) 124–136.

[19] M. Soliman, O. Malik, D. T. Westwick, Multiple model predictive control for wind turbines with doubly fed induction generators, IEEE Transactions on Sustainable Energy 2 (3) (2011) 215–225.

[20] S. Bououden, M. Chadli, S. Filali, A. El Hajjaji, Fuzzy model based multivariable predictive control of a variable speed wind turbine: Lmi approach, Renewable Energy 37 (1) (2012) 434–439.

[21] S. Bououden, S. Filali, M. Chadli, Fuzzy predictive control of a variable speed wind turbine, Energy Procedia 42 (2013) 357–366.

[22] C. Bottasso, P. Pizzinelli, C. Riboldi, L. Tasca, Lidar-enabled model predictive control of wind turbines with real-time capabilities, Renewable Energy 71 (2014) 442–452.

[23] A. Bemporad, M. Morari, V. Dua, and E.N. Pistikopoulos. The explicit linear quadratic regulator for constrained systems, Automatica 38 (2002) 3-20.

[24] F. Bayat, T.A. Johansen, and A.A. Jalali. Using hash tables to manage the time-storage complexity in a point location problem: Application to explicit model predictive control, Automatica 47 (2011) 571-577.

[25] M.A. Mohammadkhani, F. Bayat, and A.A. Jalali. Constrained linear parameter-varying control using approximate multiparametric programming, Optimal Control Applications and Methods 39 (2018) 1670-1683.

[26] K. Hooshmandi, F. Bayat, M.R. Jahed-Motlagh, and A.A. Jalali, Stability analysis and design of nonlinear sampled-data systems under aperiodic samplings, International Journal of Robust and Nonlinear Control 28 (2018) 2679-2699.

[27] K. Hooshmandi, F. Bayat, M.R. Jahed-Motlagh, and A.A. Jalali, Polynomial LPV approach to robust H∞ control of nonlinear sampled-data systems, International Journal of Control (2018), In press, DOI: 10.1080/00207179.2018.1547422.